

\magnification=1200
\vsize=8.9 true in
\hsize=6.5 true in
\overfullrule=0pt

\pretolerance=10000
\tolerance=10000
\baselineskip=12pt plus1pt minus1pt

\def\gtorder{\mathrel{\raise.3ex\hbox{$>$}\mkern-14mu
             \lower0.6ex\hbox{$\sim$}}}
\def\ltorder{\mathrel{\raise.3ex\hbox{$<$}\mkern-14mu
             \lower0.6ex\hbox{$\sim$}}}

\font\tenbi=cmmib10 \font\sevenbi=cmmib10 at 7pt
\font\fivebi=cmmib10 at 5pt
\font\tenbsy=cmbsy10\font\sevenbsy=cmbsy10 at 7pt
\font\fivebsy=cmbsy10 at 5pt

\def\boldpoint{\def\bf{\fam0\tenbf}
  \textfont0=\tenbf \scriptfont0=\sevenbf \scriptscriptfont0=\fivebf
  \textfont1=\tenbi \scriptfont1=\sevenbi \scriptscriptfont1=\fivebi
  \textfont2=\tenbsy \scriptfont2=\sevenbsy
\scriptscriptfont2=\fivebsy
  \textfont3=\tenex \scriptfont3=\tenex \scriptscriptfont3=\tenex
  \def\it{\fam\itfam\tenit}%
  \textfont\itfam=\tenit
  \def\sl{\fam\slfam\tensl}%
  \textfont\slfam=\tensl
  \def\tt{\fam\ttfam\tentt}%
  \textfont\ttfam=\tentt
  \bf}

\centerline{\bf CAN PLASMA SCATTERING MIMIC A COSMOLOGICAL
RED--SHIFT?}
\bigskip
\centerline{S.~SCHRAMM\footnote*{Present address:  IUCF, 2401 Milo B.
Sampson Lane, Bloomington, IN 47405 USA} AND S.~E.~KOONIN}
\medskip
\centerline{W. K. Kellogg Radiation Laboratory}
\centerline{California Institute of Technology, Pasadena, CA 91125}
\bigskip

\baselineskip=24pt plus1pt minus1pt

\midinsert
\narrower\narrower
\centerline{ABSTRACT}
\medskip
We investigate the extent to which the scattering of light from
plasma fluctuations can mimic a cosmological red--shift, as has been
suggested by Wolf {\it et al}.  A plasma model for the plasma
structure function results in a spectrum of scattered light very
different from that associated with a constant red--shift, implying
that the  ``Wolf effect'' cannot be involved to perturb the
cosmological distance scale.
\endinsert

In a number of articles, Wolf and collaborators have proposed that
the scattering of light by a fluctuating medium can shift its
frequency in such a way as to mimic a cosmological red--shift (Wolf
1989).  This ``Wolf effect'' has been invoked to explain certain
unusual quasar emission features (Arp 1987).  In all previous
analyses of the phenomenon, the plasma fluctuations have been assumed
to be uncorrelated in space and time.  Since any  interaction of
photons with the plasma must have a microscopic origin, it is
possible (and indeed essential) to consider whether or not the effect
survives when more realistic plasma structure functions are used.
Such an investigation is the subject of  this paper.  Unfortunately,
we find that, under plausible circumstances,  the spectrum of
scattered light differs significantly from that associated with a
cosmological red--shift.  Hence, the Wolf effect cannot be invoked to
perturb the cosmological distance scale, although it may indeed
distort line shapes.

Following the notation in Wolf  (1989), we let $S^i(\hbox{\boldpoint
$u$},\omega)$ denote the spectrum of incoming photons with frequency
$\omega$ and direction {\boldpoint $u$}  ($\hbox{\boldpoint $u$}\cdot
\hbox{\boldpoint $u$}=1$). Using standard linear response theory, the
scattering properties of the medium can be described by a scattering
kernel ${\cal K}$ depending upon the momentum {\boldpoint $q$} and
energy $q_0$ that the incoming photons deliver to the medium (see
Figure~1; we henceforth put $\hbar=c=1$).  For a final frequency
$\omega^\prime$ obtained by scattering through an angle $\theta$,
$q_0=\omega-\omega^\prime$ and $\vert \hbox{\boldpoint $q$}\vert^2=
\omega^{\prime 2}+\omega^2-2\omega\omega^\prime\cos\theta$.

The intensity  of the light  scattered into direction
$\hbox{\boldpoint $u$}^\prime$ with frequency $\omega^\prime$ is
given by
$$
S^f(\hbox{\boldpoint $u$}^\prime,\omega^\prime) =
A\, \int^{\infty}_{-\infty}\,
d\omega ~S^{i}(\hbox{\boldpoint $u$},\omega)\,
{\cal K}\left(\omega^\prime \hbox{\boldpoint $u$}^\prime-
\omega \hbox{\boldpoint $u$}, \omega^\prime - \omega,  \omega\right)
\eqno(1)$$
with $A$ an unimportant overall constant.  The scattering
kernel~${\cal K}$ depends upon both the nature and distribution of
plasma excitations, as well as the way in which  photons couple to
them.  Wolf {\it et~al.} (1989) assume a Gaussian form for the
kernel,
$$
{\cal K}(\hbox{\boldpoint $q$}, q_0) \sim e^{- \sigma_0^2 q_0^2 -
\sum_i\sigma_i^2 q_i^2 }\eqno(2)
$$
with widths $\sigma_0, \sigma_i\, (i=1,2,3)$ for the energy and
momentum transfers, respectively. However, since energy and momentum
(or space and time) correlations are independent in equation~(2), the
expression manifestly violates causality and is therefore not a
reasonable approximation to a scattering kernel.

To use a more realistic model of the scattering kernel, we consider
scattering of light off a gas of non--interacting (quasi--)particles
with mass~$m$ at a temperature~$T$. At the low momentum transfers
associated with optical photons, these  excitations are the plasmons,
in which case $m$ is the plasma frequency.  The particle distribution
is then described by  Bose statistics:
$$
n(k)={1\over e^{E_k/T}-1} \eqno(3)
$$
and the dispersion relation for the quasiparticles is
$$
E^2_k=k^2+m^2\ .\eqno(4)
$$

The basic scattering process is shown in Figure~1.  An incoming
photon scatters from a quasiparticle in the medium through an
unspecified mechanism $\Gamma$.  We need not specify the properties
of the actual coupling $\Gamma$ between photons and quasiparticles
beyond assuming that any momentum dependence is smooth over the
linewidths of interest, as is  the case for Thomson scattering.

Under these assumptions, the scattering matrix element is simply  $S
=\Gamma \,\delta^{(4)}(p^\prime-p+q)$ and the statistical ensemble of
quasiparticles generates a scattering kernel of the form
$$\vbox{\halign{\hfil $#$ & $#$\hfil\cr
{\cal K}(\hbox{\boldpoint $q$},q_0) &=\Gamma^2
\int {{d^3k}\over{2 E_k}}\,n(k)\,\int {{d^3k^\prime}\over{2
E_{k^\prime}}}
\left( \delta^{(4)}(k^\prime-k-q) \right) ^2\cr
& = \Gamma^2 {{V T}\over{(2 \pi)^4}}
\int {{d^3k}\over{2 E_k}} n(k) {{1}\over{2 E_{k+q}}}
\delta(E_k+q_0-E_{k+q})\cr
&\equiv {A\over\omega\omega^\prime}\tilde{\cal K}(q,q_0)\cr
}}\eqno(5)$$
where $V$ is the normalization volume, $\tilde{\cal K}$ is
independent of $\omega$, and $A$ is a constant.

The temperature dependence of $\tilde{\cal K}$ is best characterized
by $\mu\equiv m/T$.  We consider two limiting cases:  $\mu=1$
(relativistic) and $\mu=10$ (non--relativistic).  Figure~2 shows, for
these two cases, the scattering kernel $\tilde{\cal K}$ assuming a
mass $m$ of 1~eV.  As one can see, $\tilde{\cal K}$ is not a
gaussian, which would have elliptical contours.  Further, the kernel
is non--zero only  in the space--like cone ($q^2>q^2_0$). In the
relativistic case ($\mu=1$) the kernel is more nearly symmetric about
the $q$--axis. In the case of low temperatures ($\mu=10$) the kernel
is shifted to positive values of $q_0$, since there are few
high--energy particles present and the incoming photon generally
loses energy to the medium.

If the incident light is a single line of frequency $\omega$, the
intensity of the scattered light is given by equation~(2) as
$$
S(\omega^\prime) = {\cal K}(\omega^\prime \hbox{\boldpoint
$u$}^\prime-\omega
\hbox{\boldpoint $u$}, \omega^\prime- \omega)\eqno(6)
$$
For a concrete illustration, we fix the scattering angle $\theta$ to
be $20^o$ as used in Wolf (1989).  Figure~3 shows the resulting line
shapes for an incoming photon with a wavelength $\lambda = 500$~nm
for various values of~$\mu$. The scattering red--shifts the centroid
of the line by an amount that depends  on the quasi--particle
mass~$m$; the line is also broadened significantly (i.e., comparable
to the shift).  Figure~4a shows the width  of the scattered line as a
function of the wavelength of the incoming photon for different
values of~$\mu$. Line widths smaller than about 100~nm can be
achieved only for large values of the mass, $\mu > 50$. However, such
large values suppress the intensity of the scattered light by about a
factor of~$e^{-\mu}$. In an astronomical context this would be an
increase of some 54 magnitudes for  $\mu = 50$.  Figure~4b shows the
red--shift of the incoming light,  $z =
(\lambda^\prime-\lambda)/\lambda$ as a function of $\lambda$, the
initial wavelength  of the photon. The red--shift varies strongly
with $\lambda$, independent of $\mu$; this behavior is very
different from the constant $z$ characterizing a cosmological
red--shift.

Finally we consider the effect of an anisotropic scattering medium.
We assume an  anisotropic dispersion relation for the quasiparticle
of the form
$$
E_k^2 = (a k_x)^2 + k_y^2 + k_z^2 + m^2\eqno(7)
$$
with an asymmetry parameter $a$ multiplying the $x$--component of the
momentum.  Such an asymmetry might arise from a magnetic field, for
example. Inserting this relation into equation~(4) one obtains $$
{\cal K}(\hbox{\boldpoint $q$},q_0) =
{{\tilde{A}}\over{q\omega\omega^\prime}} \tilde{\cal K}
(\tilde{\hbox{\boldpoint $q$}},q_0)\eqno(8)
$$
with $\tilde{\hbox{\boldpoint $q$}}=(a q_x,q_y,q_z)$. Thus, the
anisotropy can be accounted for by a  simple rescaling of $q_x$ and
our previous formulae  can be used to calculate the photon scattering
from this anisotropic medium. Figure~5 shows the change of the
spectrum of the scattered radiation as the anisotropy parameter $a$
varies. Here the incident beam is again along the $z$--axis;
i.e.,~orthogonal to the direction of anisotropy ($x$--axis). The line
is shifted to larger wavelength with decreasing~$a$. Figure~6 shows
the red--shift for fixed $a=0.4$ as a function of the photon
wavelength. The scattering angle $\theta$ is fixed at $20^\circ$. The
different curves shown in the plot are obtained by varying the angle
$\theta_m$ between incident photon and the $x$--axis (in these
calculations, $\hbox{\boldpoint $u$}, \hbox{\boldpoint $u$}^\prime$,
and the $x$--axis are assumed to be coplanar; i.e.,~$q_y=0$). The
change of the red--shift with respect to the wavelength is weaker
than in the isotropic case ($a=1$), and the width of the lines,
Figure~7, is also somewhat smaller, but still quite large.  Note that
these results are obtained for a value of $\mu=50$ which again leads
to the strong overall suppression of the signal as discussed  above.

Several conclusions can be drawn from these calculations.  First,
scattering from the simplest, isotropic statistical plasma cannot
mimic a cosmological red--shift: the redshift is a strong function of
frequency and the line is broadened by an amount comparable to the
shift.  These deficiencies can be ameliorated (but not eliminated) by
requiring that the plasma have a high anisotropy (of order $10^3$ or
more) or be strongly non--relativistic ($\mu\gg1$).  However, even in
such cases, the thermal distribution of plasmons strongly suppresses
the magnitude of the scattering.  We also note that standard models
of quasars (Rees 1989) give $n_e\ltorder 10^{10}~{\rm cm}^{-3}$ and
$T\sim10^4$~K,  implying $\mu\ltorder4\times10^{-6}$, just the
opposite extreme.

Of course, our discussion cannot rule out coherent (i.e.,
nonstatistical) fluctuations in the plasma (as might be caused by a
radio source).  However, such a source would have to have a very
narrow ($\sim 1\%$) bandwidth to avoid broadening the scattered line,
a very unlikely circumstance.  Hence, we conclude that the Wolf
effect cannot be invoked to distort cosmological red--shifts.
 \bigskip

We are grateful to Profs. G.~Neugebauer and W.~Sargent for drawing
our attention to  this problem and to Profs. S.~Phinney and
R.~Blandford for comments on the manuscript.   This work was
supported in part by the National Science Foundation, Grant Nos.
PHY90--13248 and PHY91--15574.
 \vfill\eject

\centerline{REFERENCES}
\bigskip

Arp, H. 1987, {\it Quasars, Redshift and Controversies} (Berkeley,
CA: Interstellar Media).

Rees, M., Netzer, H., and Ferland, G. J. 1989, {\it Ap. J.}, {\bf
347}, 640.

Wolf, E. 1989,  {\it Phys. Rev. Lett.},  {\bf 63}, 2220;
James, D.~F.~V., Savedoff, M.~P., and Wolf, E. 1990, {\it Ap.~J.},
{\bf 359}, 67;
James, D.~F.~V. and Wolf, E. 1990, {\it Phys.~Lett.~A}, {\bf 146},
167; see also {\it Sky and Telescope}, February 1992, p. 38.

\bigskip
\bigskip
\centerline{FIGURE CAPTIONS}
\bigskip

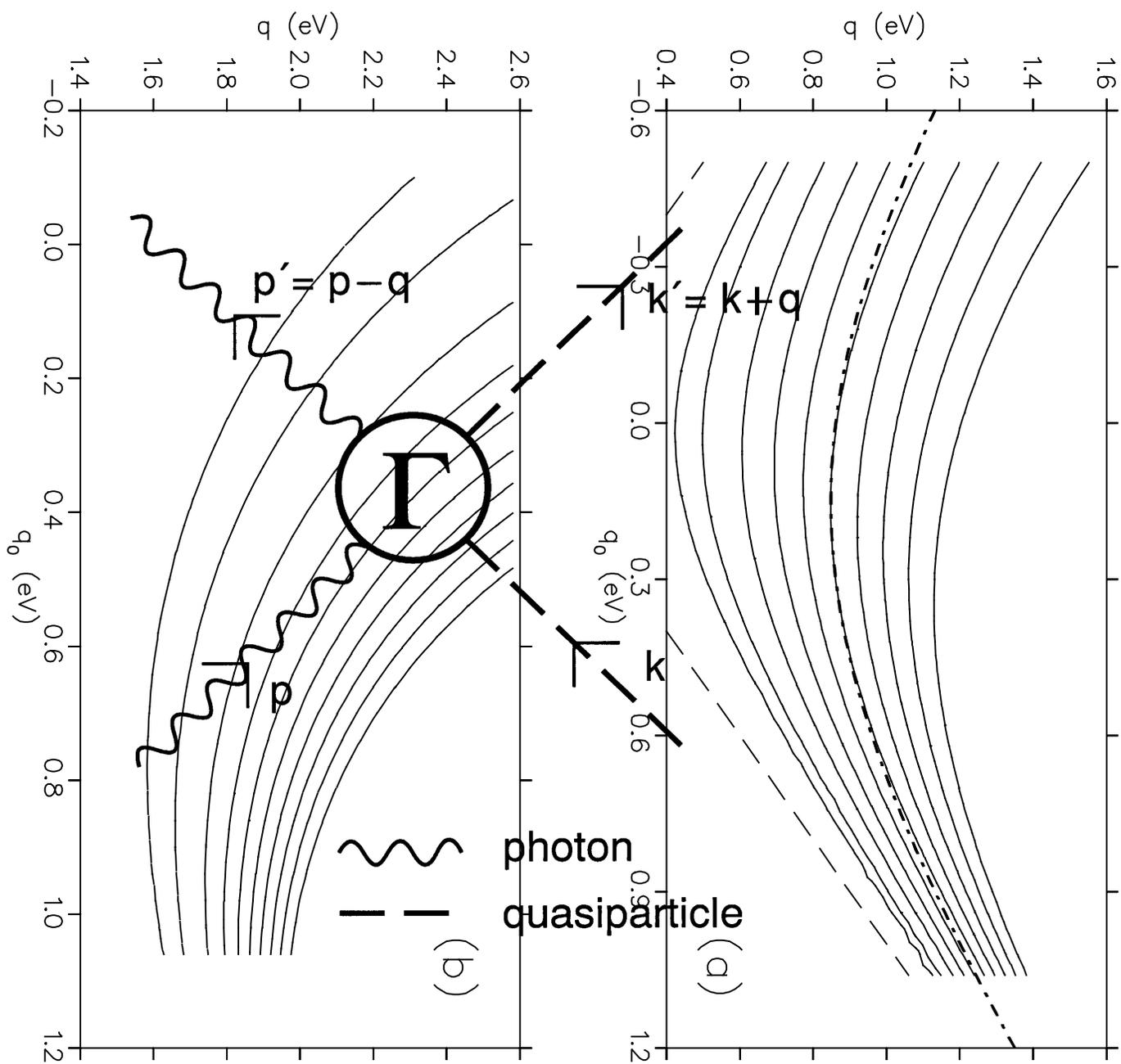
FIG.~1---Feynman graph of the photon--plasmon scattering process. The
vertex $\Gamma$ is assumed to have no significant momentum
dependence.
\bigskip

FIG.~2---Contour plot of the scattering kernel $\tilde{\cal
K}(q_0,q)$ when $\mu=m/T$ has the value (a) $\mu = 1$ and (b) $\mu =
10$, respectively.  The dashed line in (a) shows the causal limit
($q=q_0$) beyond which $\tilde{\cal K}$ must vanish.  The dash--dot
line shows the energy and momentum transfer for a photon with a
wavelength $\lambda = 500$~nm and a scattering angle $\theta =
20^\circ$.
\bigskip

FIG.~3---Spectrum of the light scattered when the initial photon has
wavelength $\lambda = 500~$nm for values of $\mu = 1, 6, \ldots,
101$. The panels (a, b, c) show the results for quasiparticle masses
$m= 0.5$, 1, and $1.5$~eV, respectively.
\bigskip

FIG.~4---(a) FWHM values and (b) red--shifts ($z$) of the scattered
line as a function of the initial photon wavelength; values of $\mu =
6,11,\ldots,101$ are shown.
\bigskip

FIG.~5---Spectrum of the scattered light when the initial line has a
wavelength of  500~nm for fixed $\mu=50$ and values of the anisotropy
parameter $a=0.4,0.5,\ldots,1$.
\bigskip

FIG.~6---FWHM values (a) and red--shifts (b) as functions of the
incident wavelength for fixed anisotropy $a=0.004$ and $\mu=50$. The
angle between the incident photon and the anisotropy axis,
$\theta_m$, varies from $50^\circ$ to $120^\circ$ in steps of
$10^\circ$.
\bigskip

 
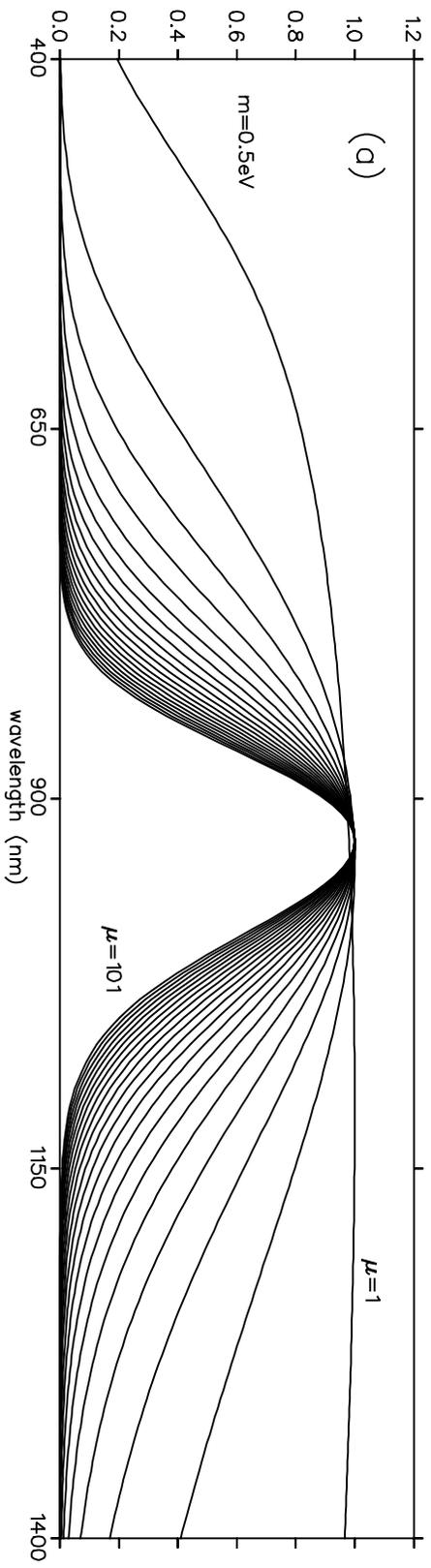
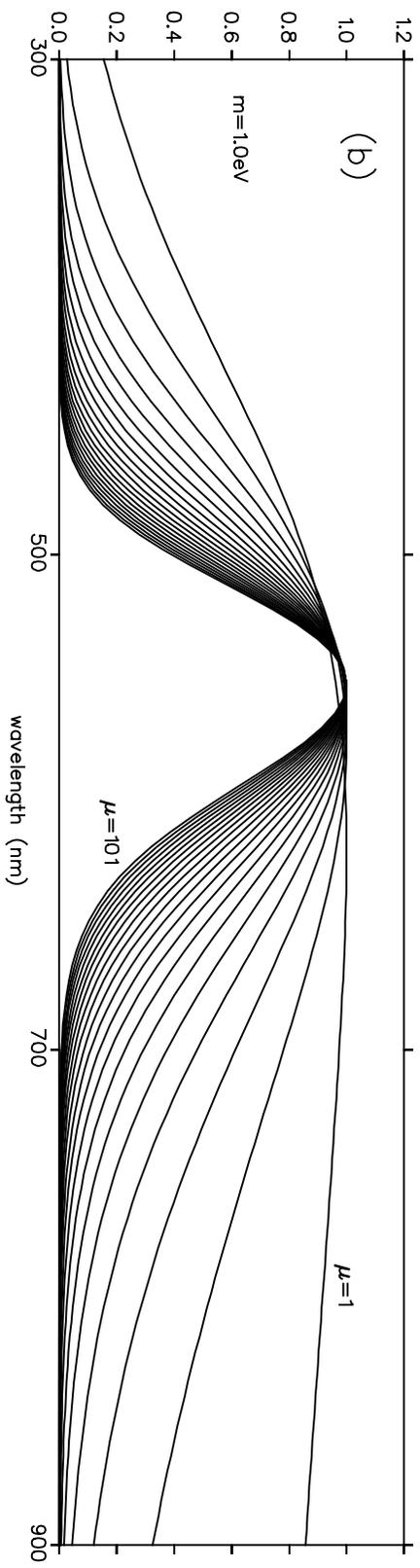
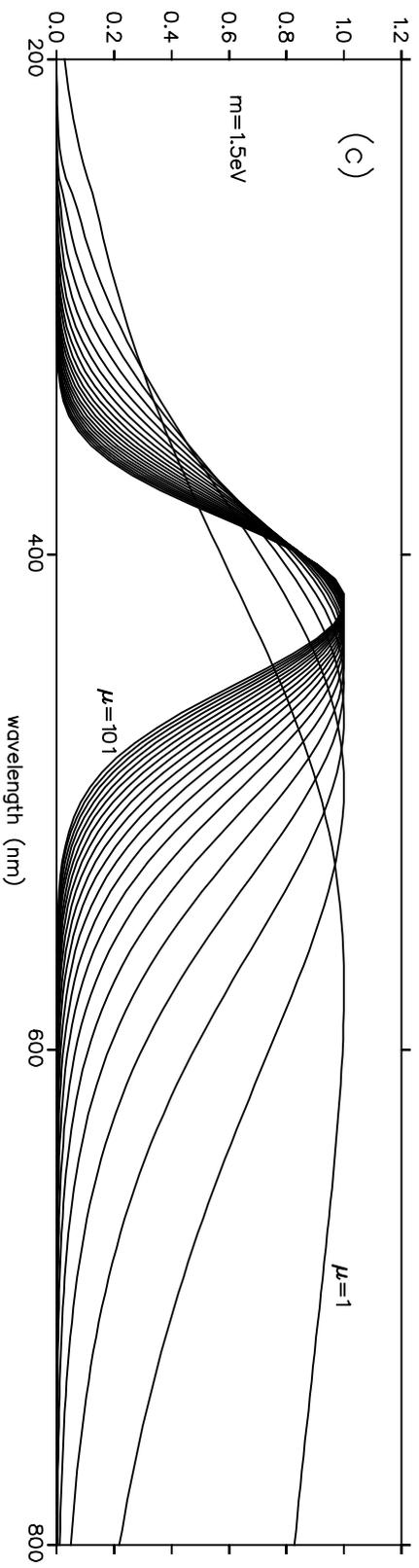
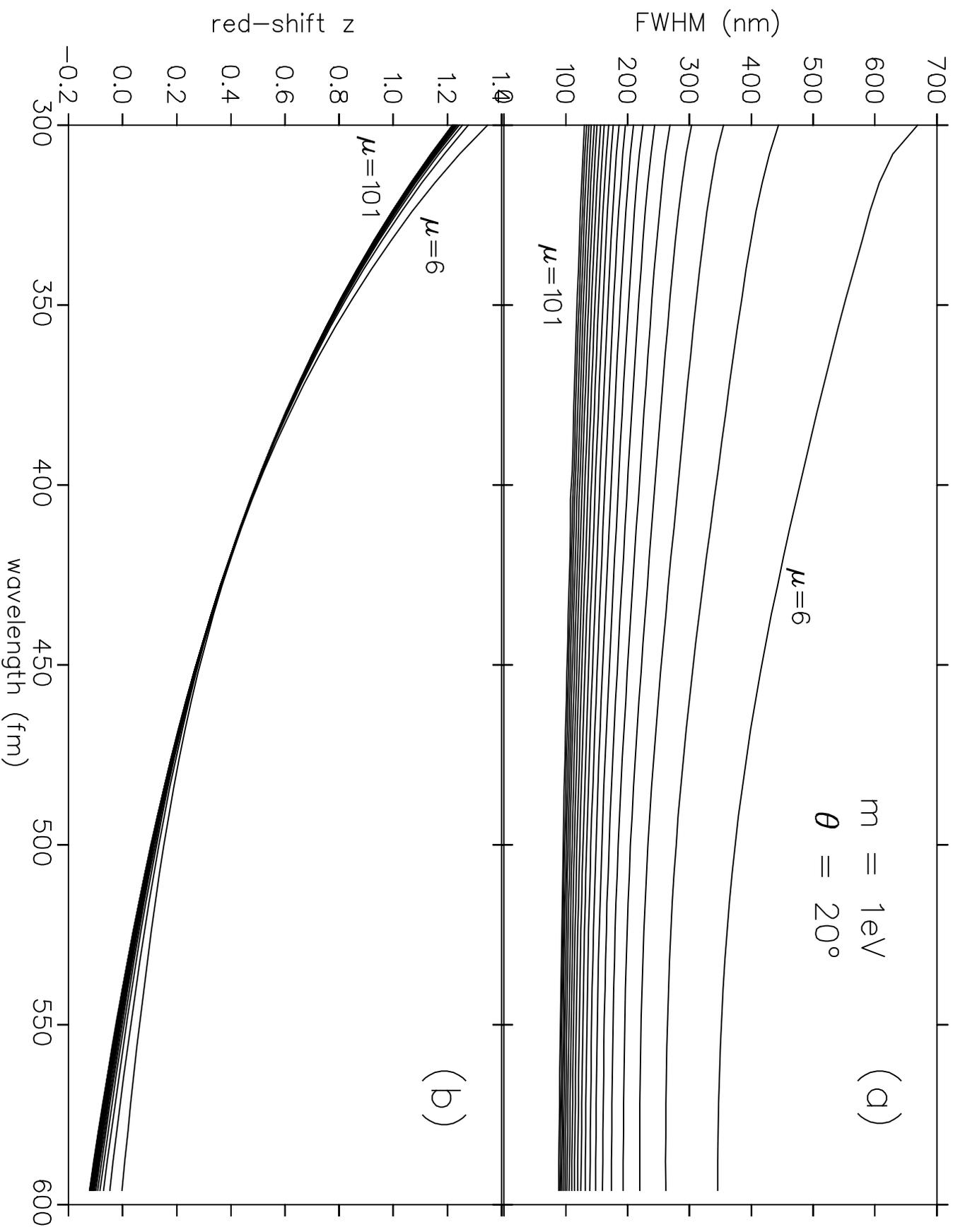
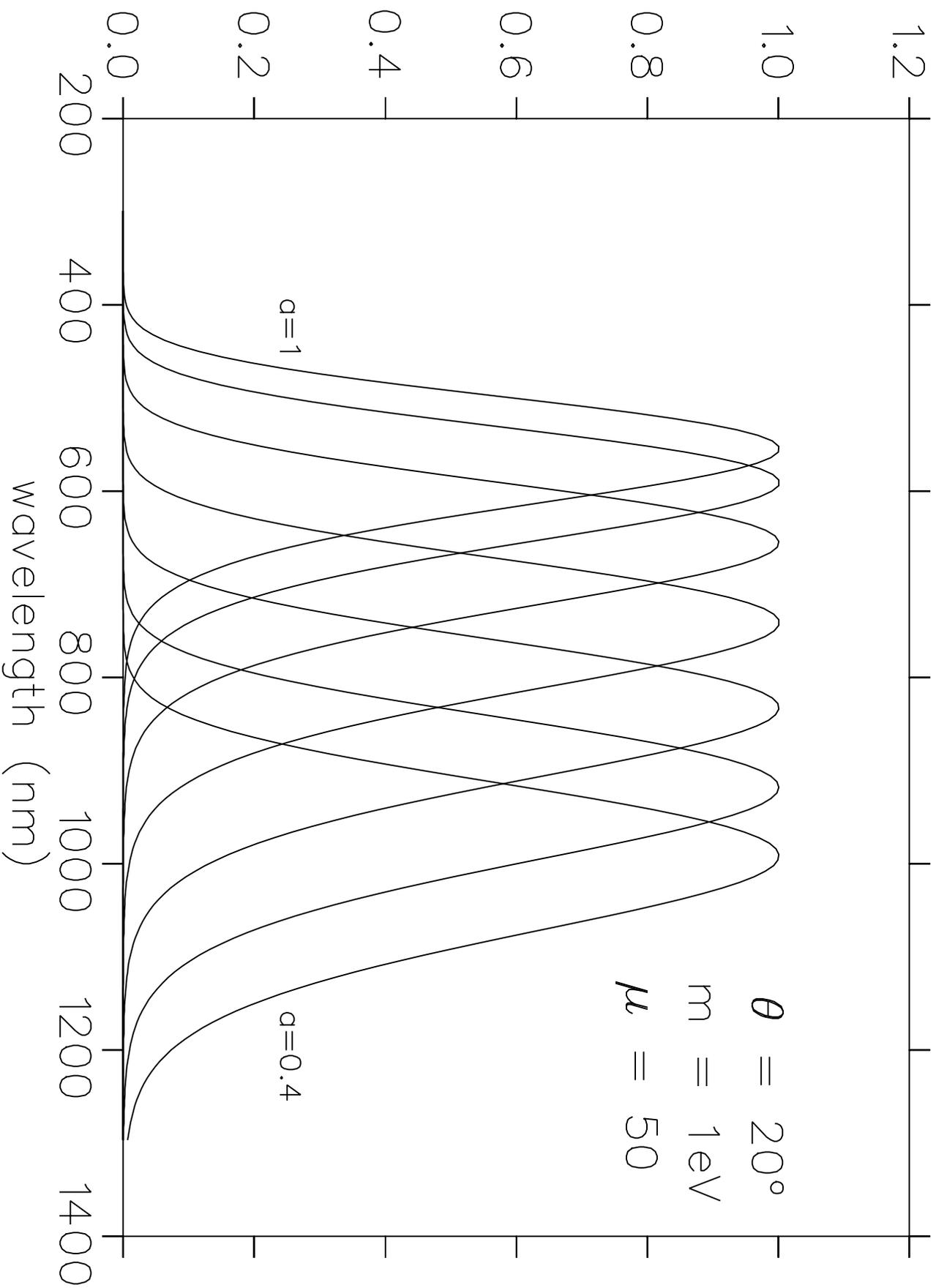
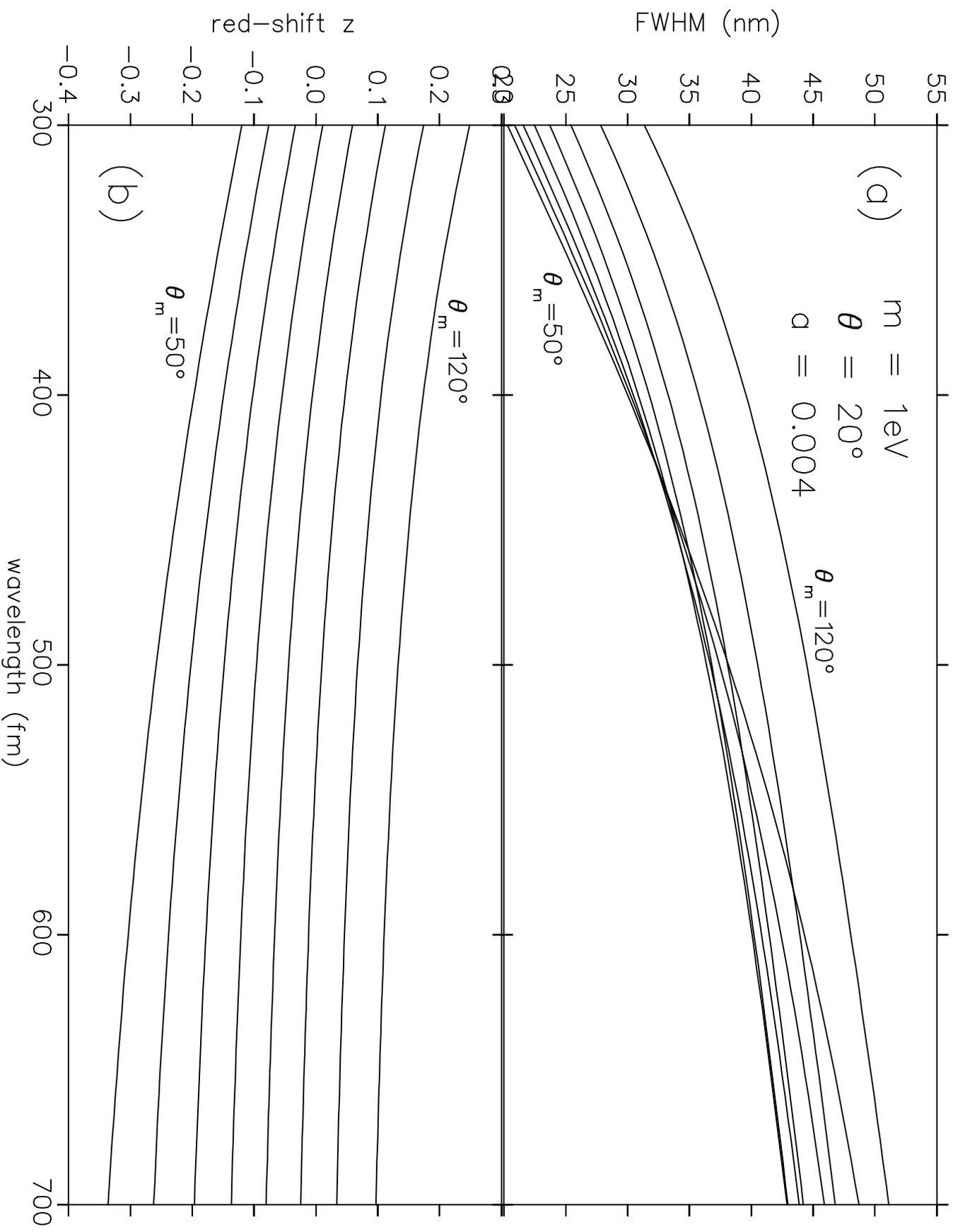
\bye